\documentclass{emulateapj}
\begin{document}

\title{The GALEX UV luminosity function of the  cluster of galaxies Abell 1367.}
\author{
L. Cortese\altaffilmark{1}, A. Boselli\altaffilmark{1},   
G. Gavazzi\altaffilmark{2}, J. Iglesias-Paramo\altaffilmark{1}, 
B. F. Madore\altaffilmark{3,4},
T. Barlow\altaffilmark{5}, L. Bianchi\altaffilmark{6},
Y.-I. Byun\altaffilmark{7}, J. Donas\altaffilmark{1},
K. Forster\altaffilmark{5}, P. G. Friedman\altaffilmark{5},
T. M. Heckman\altaffilmark{8}, P. Jelinsky\altaffilmark{9},
Y.-W. Lee\altaffilmark{7},
R. Malina\altaffilmark{1},
D. C. Martin\altaffilmark{5}, B. Milliard\altaffilmark{1},
P. Morrissey\altaffilmark{5}, S. Neff\altaffilmark{10},
R. M. Rich\altaffilmark{11}, D. Schiminovich\altaffilmark{5},
O. Siegmund\altaffilmark{9}, T. Small\altaffilmark{5},
A. S. Szalay\altaffilmark{8}, 
, M. A. Treyer\altaffilmark{5},
B. Welsh\altaffilmark{9}, T. K. Wyder\altaffilmark{5} 
}
\altaffiltext{1}{Laboratoire d'Astrophysique de Marseille, BP8, Traverse du Siphon, F-13376 Marseille, France}
\altaffiltext{2}{Universit\`a degli Studi di Milano - Bicocca, P.zza della Scienza 3,
20126 Milano, Italy}
\altaffiltext{3}{Observatories of the Carnegie Institution of Washington,
813 Santa Barbara St., Pasadena, CA 91101}
\altaffiltext{4}{NASA/IPAC Extragalactic Database, California Institute
of Technology, Mail Code 100-22, 770 S. Wilson Ave., Pasadena, CA 91125}
\altaffiltext{5}{California Institute of Technology, MC 405-47, 1200 East
California Boulevard, Pasadena, CA 91125}
\altaffiltext{6}{Center for Astrophysical Sciences, The Johns Hopkins
University, 3400 N. Charles St., Baltimore, MD 21218}
\altaffiltext{7}{Center for Space Astrophysics, Yonsei University, Seoul
120-749, Korea}
\altaffiltext{8}{Department of Physics and Astronomy, The Johns Hopkins
University, Homewood Campus, Baltimore, MD 21218}
\altaffiltext{9}{Space Sciences Laboratory, University of California at
Berkeley, 601 Campbell Hall, Berkeley, CA 94720}
\altaffiltext{10}{Laboratory for Astronomy and Solar Physics, NASA Goddard
Space Flight Center, Greenbelt, MD 20771}
\altaffiltext{11}{Department of Physics and Astronomy, University of
California, Los Angeles, CA 90095}

\begin{abstract}
We present the  GALEX NUV ($2310\rm \AA$) and FUV ($1530\rm \AA$) galaxy luminosity functions of the 
nearby cluster of galaxies A1367 in 
the magnitude range $-20.3\leq M_{AB} \leq -13.3$. 
The luminosity functions are consistent with previous ($\sim$ 2 mag shallower) estimates 
based on the FOCA and FAUST experiments, but display 
a steeper faint-end slope than the GALEX luminosity function for local field galaxies.
Using spectro-photometric optical data we select out star-forming systems from quiescent 
galaxies and study their separate contributions to the cluster luminosity 
function. We find that the UV luminosity function of cluster star-forming galaxies is consistent 
with the field. The difference between the cluster and field LF 
is entirely due to the contribution at low luminosities ($M_{AB}>-16$~mag) of non star-forming, early-type galaxies
that are significantly over dense in clusters.
\end{abstract}


\section{Introduction}
\setcounter{footnote}{0}
Recent determinations of the galaxy luminosity function (hereafter LF) at various frequencies, 
in various environments (i.e.\citealp{depropris03,madwick})
and in a number of redshift intervals (i.e.\citealp{vimos}) have improved our knowledge of galaxy evolution and the role 
played by the environment in regulating the star formation activity of galaxies. 
The optical cluster LF is significantly steeper than that in the field (\citealp{trentham}).
This steepening is due to quiescent galaxies, more frequent at low luminosities in clusters, while
the LF of cluster star forming objects is similar to that in the field (\citealp{depropris03}).
The causes of this difference might reside in the density-morphology 
relation (\citealp{dressler80,whitmore}) and in particular in the overabundance 
of dwarf ellipticals in rich clusters (\citealp{ferguson}), 
whose origin is corrently debated in the framework of the environmental effects on galaxy evolution.\\ 
The ultraviolet emission UV( $\rm \sim 2000~\AA$), being dominated by young stars of intermediate masses 
($2<M<5\rm~M_{\odot}$, \citealp{boselli}) represents an appropriate tool to identify 
and quantify star formation activity.
Before the launch of the \emph{Galaxy Evolution Explorer} (GALEX)
a few balloon-borne experiments had observed the sky at ultraviolet wavelengths \citep{smith,faust,giappo}. 
Among them, the FOCA experiment \citep{foca}
allowed the first determinations of the UV LF of local field galaxies \citep{treyer98,sullivan00} and 
of nearby clusters \citep{donas91,andreon,redshift}.
Combining the FOCA and FAUST data \cite{COGA03} determined the first composite LF of nearby clusters. 
They found 
no significant differences with the LF in the field. However this early determination was affected
by large statistical errors due to the lack 
of sufficient redshift coverage for UV selected galaxies and to the uncertainty in the 
UV background counts \citep{redshift}. 
GALEX has opened a new era of extragalactic UV astronomy. 
In particular it provides for the first time 
precise UV photometry of galaxies over large stretches of the sky \citep{xu}, thus making 
the background subtraction method more reliable than in the past.\\
In this letter we present the first determination of the UV  (NUV and FUV) LF 
in a nearby cluster of galaxies based upon GALEX measurements of galaxies brighter than $M_{AB}\sim-13.3$  that extends two magnitudes deeper than previous estimates \citep{COGA03}. 
Throughout this paper we assume for Abell 1367 a distance modulus $\rm \mu=34.84$ mag \citep{sakai00}, 
corresponding to $\rm H_{0}= 70~ km~s^{-1}~Mpc^{-1}$.

\begin{figure}
\epsscale{0.95}
\plotone{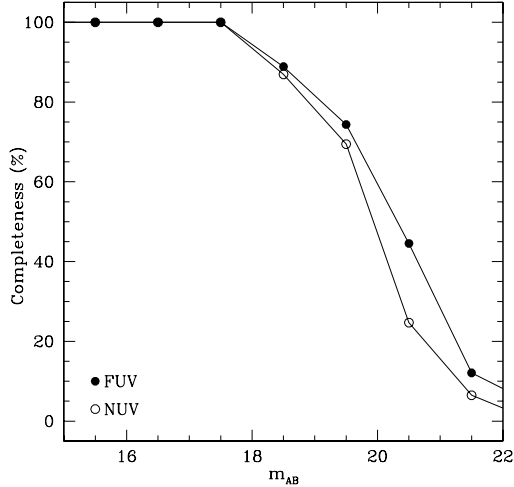}
\small{\caption{The redshift completeness per bin of UV magnitude in Abell 1367.}
\label{completeness}}
\end{figure}

\section{UV data}
GALEX provides far-ultraviolet (FUV; $\rm \lambda_{eff}=1530\AA, \Delta \lambda=400\AA$) and near-ultraviolet 
(NUV; $\rm \lambda_{eff}=2310\AA, \Delta \lambda=1000\AA$) images with a circular field of view of  $\sim$ 0.6 degrees radius.
The spatial resolution is $\sim$5 arcsec. Details of the GALEX instrument can be found in \cite{martin} 
and \cite{morrissey}.
The data analyzed in this paper consist of two GALEX pointings of the Abell cluster 1367, with a mean exposure time of 1460s, ,
centered at R.A.(J2000)=11:43:41.34 Dec(J.2000)=+20:11:24.0  (e.g. offset to the north 
of the cluster to avoid a star bright enough to threaten the detector).
Sources were detected and measured using SExtractor \citep{sex}.
The 100\% completeness limit is $m_{AB} \sim 21.5$ both in FUV and NUV \citep{xu}.
As the NUV images are
significantly deeper than the FUV, sources were selected and their parameters determined in the NUV. 
FUV parameters were extracted in the same apertures.
We used a larger SExtractor deblending parameter compared to the standard GALEX pipeline, providing reliable $\rm MAG_{AUTO}$
also for very extended sources.
The calibration uncertainty of the NUV and FUV magnitudes is $\sim 10$\% \citep{morrissey}.
Magnitudes are corrected for Galactic extinction using the \cite{schlegel98} reddening map and the Galactic extinction curve 
of \cite{cardelli89}. 
The applied extinction corrections are of 0.18 and 0.17 mag for the NUV and FUV bands respectively.
To avoid artifacts present at the edge of the field, we considered only the central 0.58 deg radius from the field center.
A reliable star/galaxy discrimination was achieved by matching
the GALEX catalog against the deepest optical catalogs available for A1367  ($B<22.5 \rm ~mag$ and $r'< 21~\rm mag$; \citealp{jorge03}), 
using a search radius of 6 arcsec, as adopted by \cite{wider} for the estimate of the GALEX local field LF.
The optical catalogs do not include a negligible part ($\sim$ 0.09 square degrees) of the GALEX field.
A total number of 292 galaxies in the FUV and of 480 galaxies in NUV with $m_{AB}\leq21.5$  are detected in the 
$\sim0.96$ square degrees field ($\sim2.5 \rm Mpc^{2}$) analyzed in this work.
 


\begin{figure*}
\epsscale{0.9}
\plottwo{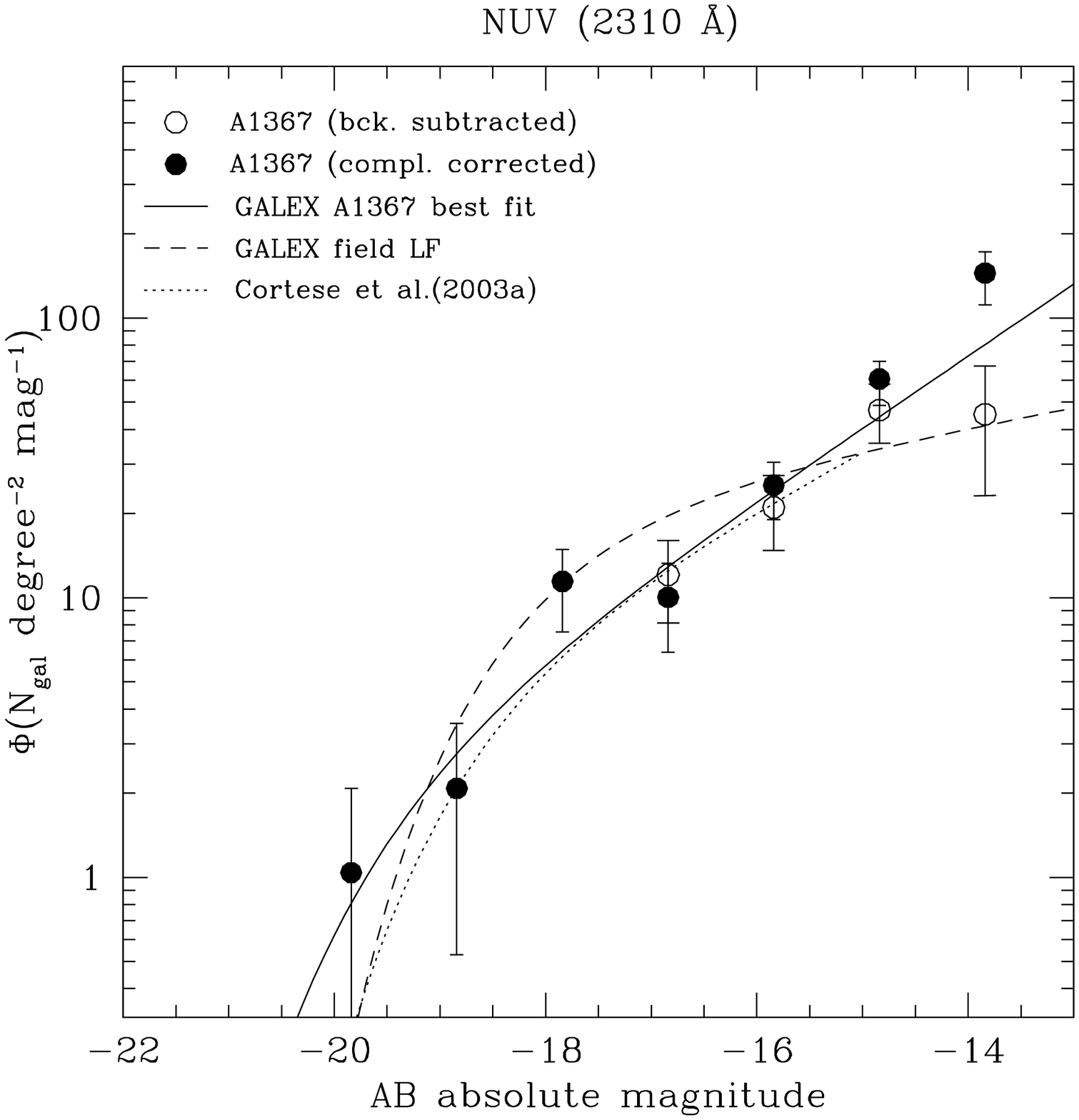}{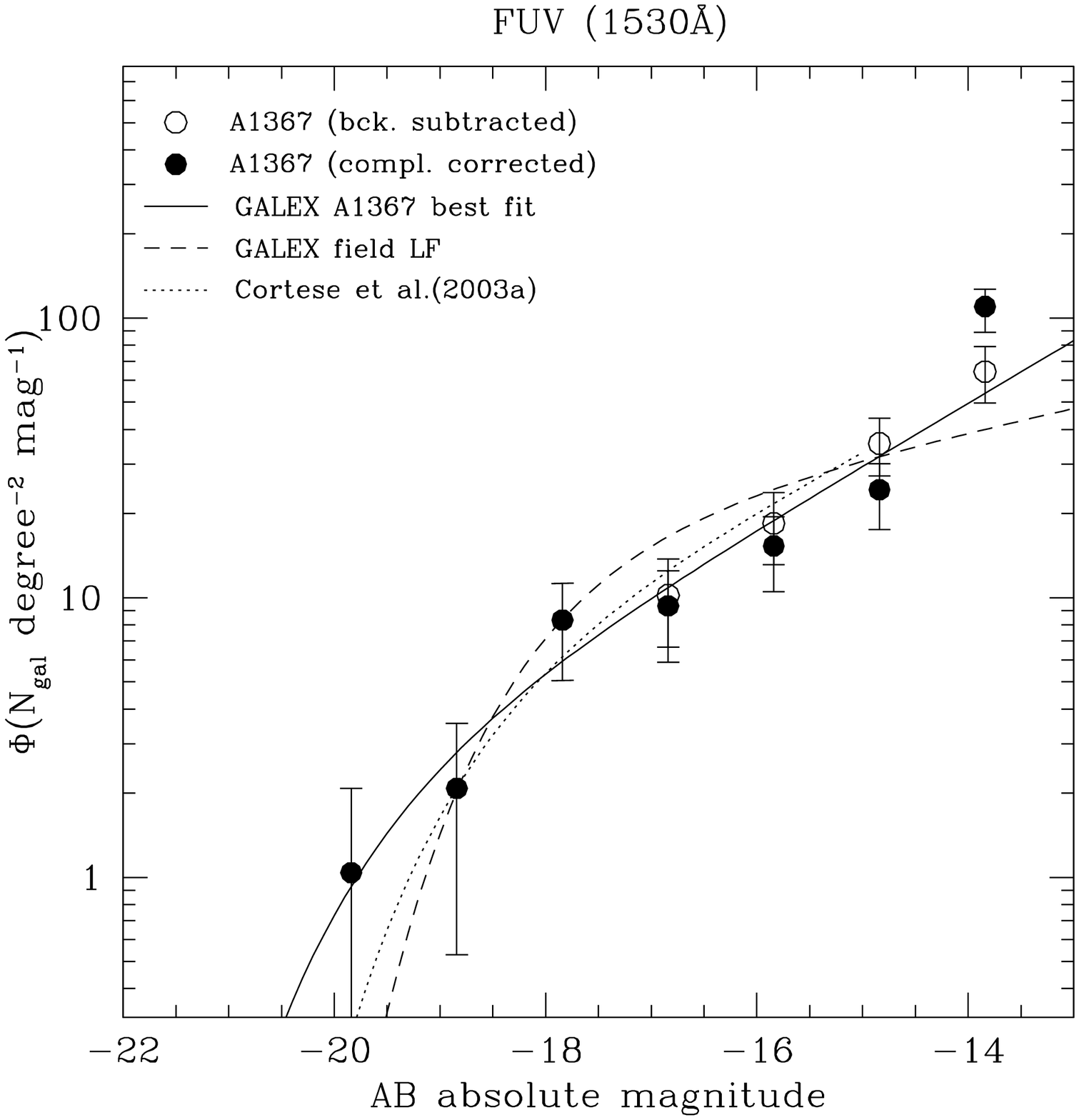}
\small{\caption{The GALEX NUV (left) and FUV (right) LF for Abell 1367. Open dots are obtained using 
the subtraction of field counts obtained by \cite{xu}; filled dots are obtained using the \emph{completeness corrected} method.
The solid line represents the best Schechter fit. The dotted line shows the composite nearby clusters 2000 $\rm \AA$ LF by \citep{COGA03}. 
The dashed line represents the GALEX local field LF \citep{wider}, normalized in order to match the cluster LF at $M_{AB} \sim -17.80$. }
\label{LFall}}
\end{figure*}

\begin{figure*}
\epsscale{0.9}
\plottwo{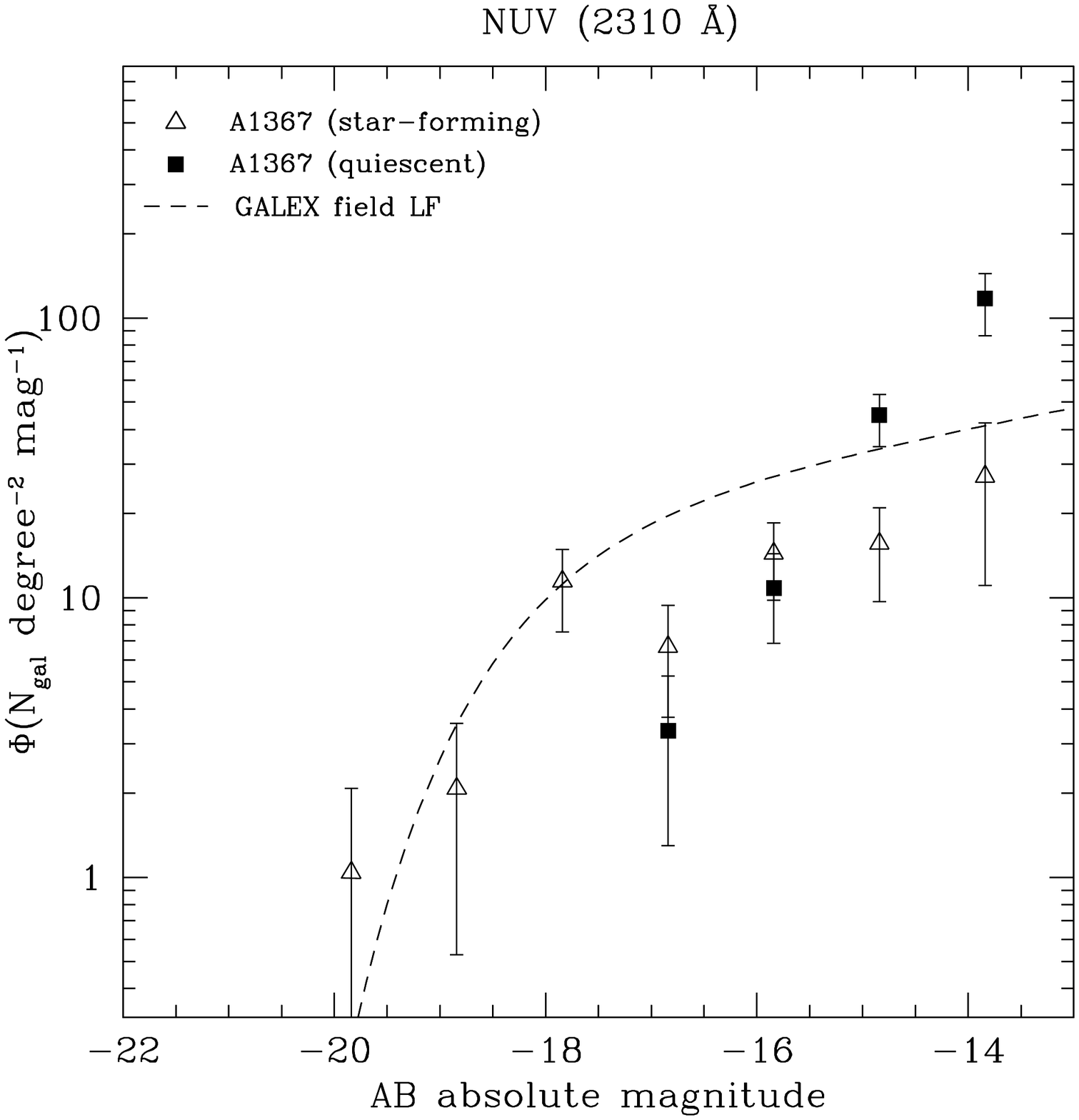}{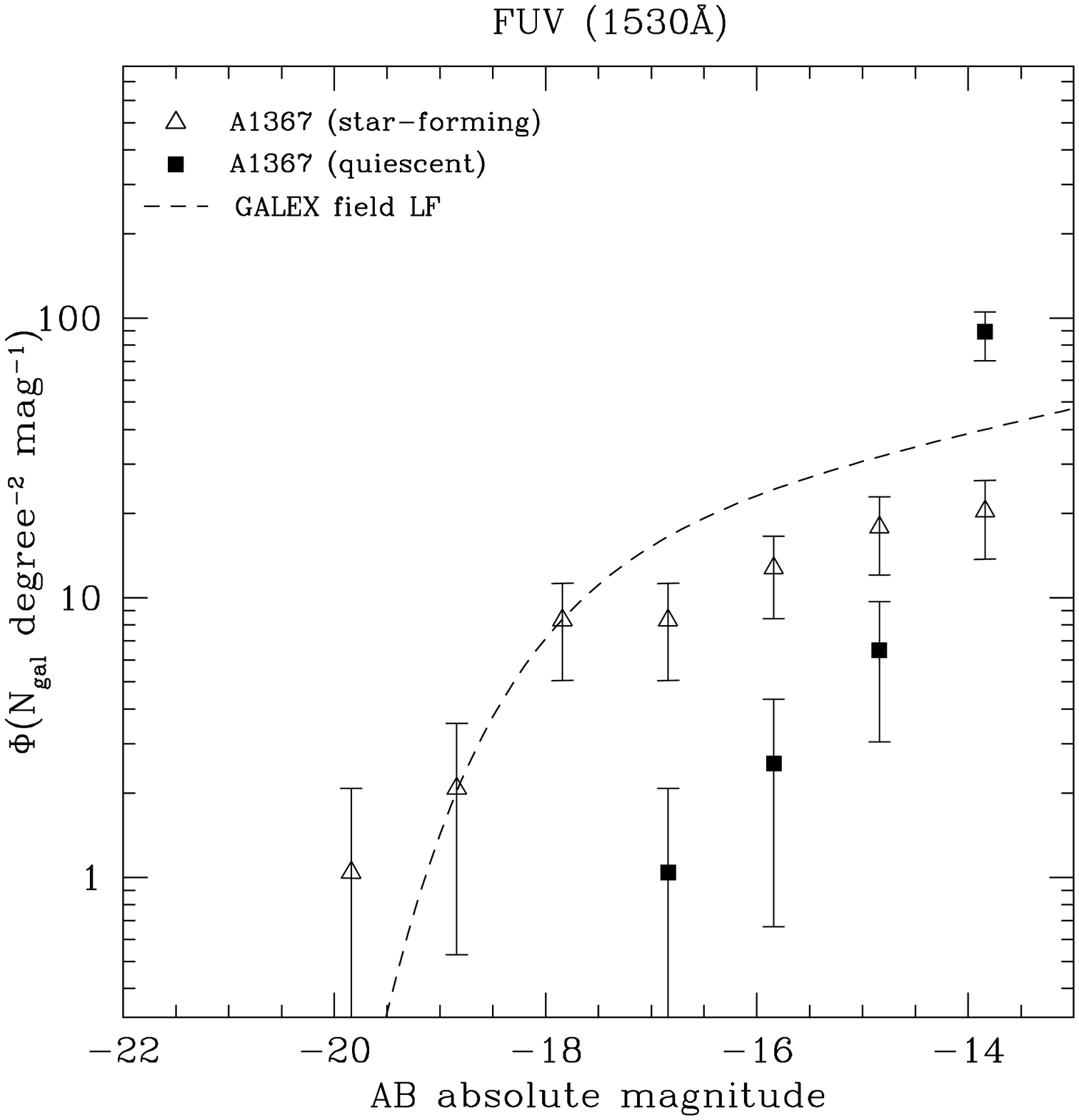}
\small{\caption{The NUV (left) and FUV (right) bi-variate LF of A1367. 
Star-forming and quiescent galaxies are indicated with empty triangles and filled squares respectively. The dashed line represents the GALEX local field 
LF \citep{wider}, normalized as in Fig.\ref{LFall}}
\label{LFtype}}
\end{figure*}

\section{The luminosity function}
The determination of the cluster LF requires a reliable estimate of the contribution 
from background/foreground objects to the UV counts. 
This can be accurately achieved for $m_{AB}\leq 18.5$, since at this limit our redshift completeness is 
$\sim 90$ \% (\citealp{redshift,COGA04}; see Fig. \ref{completeness}). 
\begin{table} 
\caption {Best Fitting Parameters.}
\label{fit}
\[
\begin{array}{p{0.22\linewidth}cccc}
\hline
\noalign{\smallskip} 
Band & Sample & \multicolumn{2}{c}{Schechter~Parameters}\\
     &        &     M^{*}    &   \alpha    \\
\noalign{\smallskip}
\hline
\noalign{\smallskip}
NUV  & A1367   & -19.77\pm0.42 & -1.64\pm0.21\\
NUV  & Field       & -18.23\pm0.11 & -1.16\pm0.07\\
\hline
FUV  & A1367   & -19.86\pm0.50 & -1.56\pm0.19	\\
FUV  & Field       & -18.04\pm0.11 &-1.22\pm0.07\\
\hline
UV($2000\rm \AA$) & Composite~cluster &  -18.79\pm0.40 & -1.50\pm0.10 \\
\noalign{\smallskip}
\hline
\end{array}
\]
\end{table}

The redshift completeness drops rapidly at magnitudes fainter than $m_{AB}\sim 18.5$, 
thus requiring the contamination 
to be estimated statistically. Two methods 
are usually applied for the computation of cluster LFs. 
The first one is based on the statistical subtraction of field galaxies, per bin of UV magnitude, that are expected to be 
randomly projected onto the cluster area, as derived by \cite{xu}. 
Alternatively, the \emph{completeness corrected} method proposed by \cite{depropris03} is to be preferred when the 
field counts have large uncertainties.  
It is based on the assumption that the UV spectroscopic sample (e.g. membership confirmed spectroscopically) 
is 'representative' of the entire cluster, 
i.e. the fraction of galaxies that are 
cluster members is the same in the (incomplete) spectroscopic sample as in the (complete) photometric one.
For each magnitude bin $i$ we count the number of cluster members $N_M$ (i.e. galaxies with velocity in the range
4000$<$V$<$10000$\rm ~km~s^{-1}$; \citealp{COGA04}), the number of galaxies with a measured recessional 
velocity $N_Z$ and the total number of galaxies $N_T$. The ratio  $N_Z$/$N_T$, corresponding to the redshift completeness in each magnitude bin 
is shown in Fig.\ref{completeness}.
The completeness-corrected number of cluster members in each bin is
$ N_i = (N_M \times N_T)/N_Z$.\\
$N_T$ is a Poisson variable, and $N_M$ is a binomial variable (the number of successes in $N_Z$ trials with probability $N_M$/$N_Z$).
Therefore the errors associated with $N_i$ are given by
$(\delta^{2}N_i/N_i^{2}) = (1/N_T) + (1/N_M) - (1/N_Z)$.
The NUV and FUV LFs using both methods (see Fig \ref{LFall}) are in good agreement for $M_{AB}\geq-14.3$.
In the last bin the two methods are inconsistent as the \emph{completeness corrected} method predicts 
a higher slope 
than the statistical background subtraction.
This disagreement is likely due to the severe redshift incompleteness for $M_{AB}\geq-14.3$. 
In any case we take the weighted mean of the two determinations.\\
Due to the small number of galaxies populating the high luminosity bins 
(i.e. only three objects brighter than $M_{AB}\sim-18.3$), 
the LFs are not well fitted with a Schechter function \citep{schechter}: the best-fit $M_{*}$ turns 
out to be brighter than the brightest 
observed galaxy.For this reason we first determine the faint-end ($-18.3\leq M_{AB}\leq-13.3$) slope in each band, fitting the LFs 
with a power law  ($\Phi(M) = c~10^{k M}$) by minimizing $\chi^{2}$.
The $\alpha$ parameter of the Schechter function can be derived from 
$k$ using the relation $\alpha = - (k/0.4 + 1)$.
Then we fit the LFs with a Schechter function, keeping $\alpha$ fixed to the value previously obtained.
This is not the canonical Schechter fit, but it provides a more realistic set of parameters than  
using a three-free-parameter fit.
The best fit parameters and their errors are listed in Table.\ref{fit}.\\
In order to  separate the contribution to the LF of star-forming from quiescent galaxies, we divide the sample 
into two classes. Using H$\alpha$ imaging data \citep{jorge,catinella,haanna,goldmine} and 
optical spectroscopy \citep{redshift,COGA04}
we can discriminate between star-forming ($EW(H\alpha)>0~\rm \AA$) and quiescent ($EW(H\alpha)=0~\rm \AA$) objects.
Unfortunately neither UV field counts for different morphological types nor a measure of $EW(H\alpha)$ for all the 
UV selected galaxies are available.
Thus we can only apply the \emph{completeness corrected} method to determine the bi-variate LFs. 
We assume that in each bin of magnitude the fraction of star-forming and quiescent cluster members 
is the same in the (incomplete) spectroscopic sample as in the (complete) photometric 
sample. The bi-variate LFs derived by this method are shown in Fig.\ref{LFtype}. 
\begin{figure}
\epsscale{1.03}
\plotone{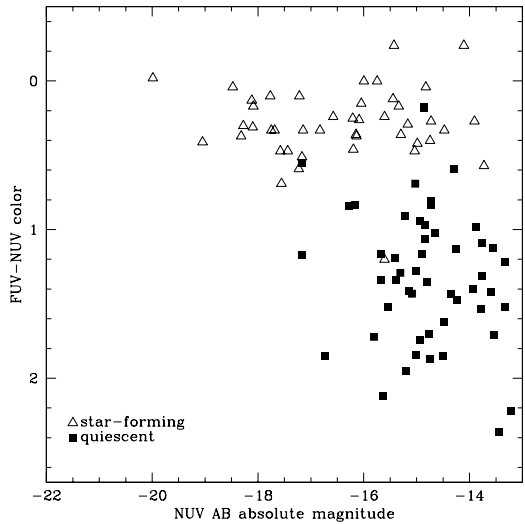}
\small{\caption{The FUV-NUV color magnitude relation for confirmed members of A1367. Symbols are as in Fig.\ref{LFtype}}
\label{colmag}}
\end{figure}

\section{Discussion}
As shown in Fig.\ref{LFall}, the GALEX LFs have a shape consistent to the composite LF 
of nearby clusters as constructed by \cite{COGA03}. Conversely, whatever fitting procedure one adopts, 
they show a steeper faint-end 
slope and a brighter $M^{*}$ than the GALEX field LF recently determined by \cite{wider}.
The brighter $M^{*}$ 
observed in Abell1367 is probably to be ascribed to the particular galaxy population of this cluster.
In fact Abell 1367 is a young cluster of galaxies composed  of at least four dynamical units at the early 
stage of a multiple merging event \citep{COGA04}.
Some galaxies have their star formation enhanced due to the interaction 
with the cluster environment, and it is this population that is responsible for the bright $M^{*}$ observed in this cluster.\\
Conversely the high faint-end slope observed in this cluster is due to the significant 
contribution of non star-forming systems at faint UV magnitudes.
In fact, as shown in Fig.\ref{LFtype}, star-forming galaxies dominate the 
UV LF for $M_{AB}\leq -17$ mag, as \cite{donas91} 
concluded for the first time.
For $M_{AB}\geq -16$ mag however, the number of red galaxies 
increases very rapidly\footnote{The bi-variate LFs cannot be compared with 
the ones computed by \cite{treyer04} for the field, since their samples do not contain ellipticals 
but only spiral galaxies.}.
This result is consistent with an UV LF constructed starting from the $r'$ LF computed by \cite{jorge03}: 
assuming a mean color $NUV-r'\sim1$ mag and $NUV-r'\sim5$  for star-forming and quiescent galaxies 
respectively, 
we are able to reproduce the contribution, at low UV luminosities, of elliptical galaxies.  
We conclude that a significant fraction of  
the low luminosity UV emission comes from extreme Horizontal Branch stars and their 
progenitors \citep{connell}.
This conclusion is reinforced by the FUV-NUV color magnitude relation 
(computed only for confirmed cluster members) shown in Fig.\ref{colmag}.
The star-forming objects dominate at high UV luminosities while the quiescent systems 
contribute more at faint magnitudes. 
Their mean
FUV-NUV color is $\sim1.5$ mag thus they influence the LF at higher 
luminosities in the NUV than in the FUV (see Fig.\ref{LFtype}).
Finally, the LFs of cluster star-forming systems have a faint-end slope 
($ \alpha \rm \sim -1.25\pm 0.2$) consistent within the statistical uncertainties with 
the GALEX field LF.
This result is expected since in the field the fraction of quiescent systems is significantly lower than that of
star forming objects \citep{dressler80,whitmore}, thus their contribution to the LF is negligible.

\acknowledgements
GALEX (Galaxy Evolution Explorer) is a NASA Small Explorer, launched in April 2003.
We gratefully acknowledge NASA's support for construction, operation,
and science analysis for the GALEX mission,
developed in cooperation with the Centre National d'Etudes Spatiales
of France and the Korean Ministry of Science and Technology. 
This research has made extensive use of the GOLDMine Database.


\begin{thebibliography}{33}
\expandafter\ifx\csname natexlab\endcsname\relax\def\natexlab#1{#1}\fi

\bibitem[{{Andreon}(1999)}]{andreon}
{Andreon}, S. 1999, \aap, 351, 65

\bibitem[{{Bertin} \& {Arnouts}(1996)}]{sex}
{Bertin}, E. \& {Arnouts}, S. 1996, \aaps, 117, 393

\bibitem[{{Boselli} {et~al.}(2001){Boselli}, {Gavazzi}, {Donas}, \&
  {Scodeggio}}]{boselli}
{Boselli}, A., {Gavazzi}, G., {Donas}, J., \& {Scodeggio}, M. 2001, \aj, 121,
  753

\bibitem[{{Cardelli} {et~al.}(1989){Cardelli}, {Clayton}, \&
  {Mathis}}]{cardelli89}
{Cardelli}, J.~A., {Clayton}, G.~C., \& {Mathis}, J.~S. 1989, \apj, 345, 245

\bibitem[{{Cortese} {et~al.}(2004){Cortese}, {Gavazzi}, {Boselli},
  {Iglesias-Paramo}, \& {Carrasco}}]{COGA04}
{Cortese}, L., {Gavazzi}, G., {Boselli}, A., {Iglesias-Paramo}, J., \&
  {Carrasco}, L. 2004, \aap, 425, 429

\bibitem[{{Cortese} {et~al.}(2003{\natexlab{a}}){Cortese}, {Gavazzi},
  {Boselli}, {Iglesias-Paramo}, {Donas}, \& {Milliard}}]{COGA03}
{Cortese}, L., {Gavazzi}, G., {Boselli}, A., {et~al.} 2003{\natexlab{a}}, \aap,
  410, L25

\bibitem[{{Cortese} {et~al.}(2003{\natexlab{b}}){Cortese}, {Gavazzi},
  {Iglesias-Paramo}, {Boselli}, \& {Carrasco}}]{redshift}
{Cortese}, L., {Gavazzi}, G., {Iglesias-Paramo}, J., {Boselli}, A., \&
  {Carrasco}, L. 2003{\natexlab{b}}, \aap, 401, 471

\bibitem[{{De Propris} {et~al.}(2003){De Propris}, {Colless}, {Driver},
  {Couch}, {Peacock}, {Baldry}, {Baugh}, {Bland-Hawthorn}, {Bridges}, {Cannon},
  {Cole}, {Collins}, {Cross}, {Dalton}, {Efstathiou}, {Ellis}, {Frenk},
  {Glazebrook}, {Hawkins}, {Jackson}, {Lahav}, {Lewis}, {Lumsden}, {Maddox},
  {Madgwick}, {Norberg}, {Percival}, {Peterson}, {Sutherland}, \&
  {Taylor}}]{depropris03}
{De Propris}, R., {Colless}, M., {Driver}, S.~P., {et~al.} 2003, \mnras, 342,
  725

\bibitem[{{Donas} {et~al.}(1991){Donas}, {Milliard}, \& {Laget}}]{donas91}
{Donas}, J., {Milliard}, B., \& {Laget}, M. 1991, \aap, 252, 487

\bibitem[{{Dressler}(1980)}]{dressler80}
{Dressler}, A. 1980, \apj, 236, 351

\bibitem[{{Ferguson} \& {Sandage} (1991){Ferguson},  \& {Sandage}}]{ferguson}
{Ferguson}, H.~C. \& {Sandage}, A. 1991, \aj, 101, 765

\bibitem[{{Gavazzi} {et~al.}(2003){Gavazzi}, {Boselli}, {Donati}, {Franzetti},
  \& {Scodeggio}}]{goldmine}
{Gavazzi}, G., {Boselli}, A., {Donati}, A., {Franzetti}, P., \& {Scodeggio}, M.
  2003, \aap, 400, 451

\bibitem[{{Gavazzi} {et~al.}(2002){Gavazzi}, {Boselli}, {Pedotti}, {Gallazzi},
  \& {Carrasco}}]{haanna}
{Gavazzi}, G., {Boselli}, A., {Pedotti}, P., {Gallazzi}, A., \& {Carrasco}, L.
  2002, \aap, 386, 114

\bibitem[{{Gavazzi} {et~al.}(1998){Gavazzi}, {Catinella}, {Carrasco},
  {Boselli}, \& {Contursi}}]{catinella}
{Gavazzi}, G., {Catinella}, B., {Carrasco}, L., {Boselli}, A., \& {Contursi},
  A. 1998, \aj, 115, 1745

\bibitem[{{Iglesias-P{\' a}ramo} {et~al.}(2002){Iglesias-P{\' a}ramo},
  {Boselli}, {Cortese}, {V{\'{\i}}lchez}, \& {Gavazzi}}]{jorge}
{Iglesias-P{\' a}ramo}, J., {Boselli}, A., {Cortese}, L., {V{\'{\i}}lchez},
  J.~M., \& {Gavazzi}, G. 2002, \aap, 384, 383

\bibitem[{{Iglesias-P{\' a}ramo} {et~al.}(2003){Iglesias-P{\' a}ramo},
  {Boselli}, {Gavazzi}, {Cortese}, \& {V{\'{\i}}lchez}}]{jorge03}
{Iglesias-P{\' a}ramo}, J., {Boselli}, A., {Gavazzi}, G., {Cortese}, L., \&
  {V{\'{\i}}lchez}, J.~M. 2003, \aap, 397, 421

\bibitem[{{Ilbert} {et~al.}(2004){Ilbert}, {Tresse}, {Zucca}, {Bardelli},
  {Arnouts}, {Zamorani}, {Pozzetti}, {Bottini}, {Garilli}, {LeBrun}, {F{\`
  e}vre}, {Maccagni}, {Picat}, {Scaramella}, {Scodeggio}, {Vettolani},
  {Zanichelli}, {Adami}, {Arnaboldi}, {Bolzonella}, {Cappi}, {Charlot},
  {Contini}, {Foucaud}, {Franzetti}, {Gavignaud}, {Guzzo}, {Iovino},
  {McCracken}, {Marano}, {Marinoni}, {Mathez}, {Mazure}, {Meneux}, {Merighi},
  {Paltani}, {Pello}, {Pollo}, {Radovich}, {Bondi}, {Bongiorno}, {Busarello},
  {Ciliegi}, {Mellier}, {Merluzzi}, {Ripepi}, \& {Rizzo}}]{vimos}
{Ilbert}, O., {Tresse}, L., {Zucca}, E., {et~al.} 2004, astro-ph/0409134

\bibitem[{{Kodaira} {et~al.}(1990){Kodaira}, {Watanabe}, {Onaka}, \&
  {Tanaka}}]{giappo}
{Kodaira}, K., {Watanabe}, T., {Onaka}, T., \& {Tanaka}, W. 1990, \apj, 363,
  422

\bibitem[{{Lampton} {et~al.}(1990){Lampton}, {Deharveng}, \& {Bowyer}}]{faust}
{Lampton}, M., {Deharveng}, J.~M., \& {Bowyer}, S. 1990, in IAU Symp. 139, 449

\bibitem[{{Madgwick} {et~al.}(2002){Madgwick}, {Lahav}, {Baldry}, {Baugh},
  {Bland-Hawthorn}, {Bridges}, {Cannon}, {Cole}, {Colless}, {Collins}, {Couch},
  {Dalton}, {De Propris}, {Driver}, {Efstathiou}, {Ellis}, {Frenk},
  {Glazebrook}, {Jackson}, {Lewis}, {Lumsden}, {Maddox}, {Norberg}, {Peacock},
  {Peterson}, {Sutherland}, \& {Taylor}}]{madwick}
{Madgwick}, D.~S., {Lahav}, O., {Baldry}, I.~K., {et~al.} 2002, \mnras, 333,
  133

\bibitem[{{Martin} {et~al.}(2005){Martin}, {Fanson}, {Schiminovich},
  {Morrissey}, {Firedman}, \& {Barlow}}]{martin}
{Martin}, D.~C., {Fanson}, J., {Schiminovich}, D., {et~al.} 2005, ApJL, Galex
  special issue (astro-ph/0411302)

\bibitem[{{Milliard} {et~al.}(1991){Milliard}, {Donas}, \& {Laget}}]{foca}
{Milliard}, B., {Donas}, J., \& {Laget}, M. 1991, Advances in Space Research,
  11, 135

\bibitem[{{Morrissey} {et~al.}(2005){Morrissey}, {Schiminovich}, {Barlow},
  {Martin}, {Blakkolb}, \& {Conrow}}]{morrissey}
{Morrissey}, P., {Schiminovich}, D., {Barlow}, T.~A., {et~al.} 2005, ApJL,
  Galex special issue (astro-ph/0411310)

\bibitem[{{O'Connell}(1999)}]{connell}
{O'Connell}, R.~W. 1999, \araa, 37, 603

\bibitem[{{Sakai} {et~al.}(2000){Sakai}, {Mould}, {Hughes}, {Huchra}, {Macri},
  {Kennicutt}, {Gibson}, {Ferrarese}, {Freedman}, {Han}, {Ford}, {Graham},
  {Illingworth}, {Kelson}, {Madore}, {Sebo}, {Silbermann}, \&
  {Stetson}}]{sakai00}
{Sakai}, S., {Mould}, J.~R., {Hughes}, S.~M.~G., {et~al.} 2000, \apj, 529, 698

\bibitem[{{Schechter}(1976)}]{schechter}
{Schechter}, P. 1976, \apj, 203, 297

\bibitem[{{Schlegel} {et~al.}(1998){Schlegel}, {Finkbeiner}, \&
  {Davis}}]{schlegel98}
{Schlegel}, D.~J., {Finkbeiner}, D.~P., \& {Davis}, M. 1998, \apj, 500, 525

\bibitem[{{Smith} \& {Cornett}(1982)}]{smith}
{Smith}, A.~M. \& {Cornett}, R.~H. 1982, \apj, 261, 1

\bibitem[{{Sullivan} {et~al.}(2000){Sullivan}, {Treyer}, {Ellis}, {Bridges},
  {Milliard}, \& {Donas}}]{sullivan00}
{Sullivan}, M., {Treyer}, M.~A., {Ellis}, R.~S., {et~al.} 2000, \mnras, 312,
  442
\bibitem[{{Trentham} {et~al.}(2005)}]{trentham}
{Trentham}, N., {Sampson}, L. \& {Banerji}, M. 2005, \mnras, in press (astro-ph/0412124)

\bibitem[{{Treyer} {et~al.}(2005){Treyer}, {Wyder}, {Schiminovich}, {Arnouts},
  {Budavari}, \& {Milliard}}]{treyer04}
{Treyer}, M., {Wyder}, T.~K., {Schiminovich}, D., {et~al.} 2005, ApJL, Galex
  special issue (astro-ph/0411308)

\bibitem[{{Treyer} {et~al.}(1998){Treyer}, {Ellis}, {Milliard}, {Donas}, \&
  {Bridges}}]{treyer98}
{Treyer}, M.~A., {Ellis}, R.~S., {Milliard}, B., {Donas}, J., \& {Bridges},
  T.~J. 1998, \mnras, 300, 303

\bibitem[{{Whitmore} {et~al.}(1993){Whitmore}, {Gilmore}, \&
  {Jones}}]{whitmore}
{Whitmore}, B.~C., {Gilmore}, D.~M., \& {Jones}, C. 1993, \apj, 407, 489

\bibitem[{{Wyder} {et~al.}(2005){Wyder}, {Treyer}, {Milliard}, {Schiminovich},
  {Arnouts}, {Budav{\' a}ri}, {Barlow}, {Bianchi}, {Byun}, {Donas}, {Forster},
  {Friedman}, {Heckman}, {Jelinsky}, {Lee}, {Madore}, {Malina}, {Martin},
  {Morrissey}, {Neff}, {Rich}, {Siegmund}, {Small}, {Szalay}, \&
  {Welsh}}]{wider}
{Wyder}, T.~K., {Treyer}, M.~A., {Milliard}, B., {et~al.} 2005, ApJL, Galex
  special issue (astro-ph/0411364)

\bibitem[{{Xu} {et~al.}(2005){Xu}, {Donas}, {Arnouts}, {Seibert},
  {Iglesias-Paramo}, \& {Blaizot}}]{xu}
{Xu}, C.~G., {Donas}, J., {Arnouts}, S., {et~al.} 2005, ApJL, Galex special
  issue (astro-ph/0411317)

\end{thebibliography}
\end{document}